\date{}
\documentclass[10pt,aps,prl,twocolumn,superscriptaddress]{revtex4-1}
\usepackage{graphicx,psfrag,amsmath,amssymb,amsfonts,color}
\usepackage[T1]{fontenc}
\usepackage{natbib}
\usepackage[hidelinks]{hyperref}
\usepackage{upgreek}
\usepackage{color}
\usepackage{ulem}
\usepackage[nice]{nicefrac}

\newcommand{\beq}{\begin{equation}}
\newcommand{\eeq}{\end{equation}}
\newcommand{\bea}{\begin{eqnarray}}
\newcommand{\eea}{\end{eqnarray}}

\begin{document}

\title{Observation of Space-Dependent Rotational Doppler Shifts with a Single Ion Probe}

\author{Nicolás A. Nuñez Barreto}
\affiliation{Universidad de Buenos Aires, Facultad de Ciencias Exactas y Naturales, Departamento de Física. Buenos Aires, Argentina.}
\affiliation{CONICET - Universidad de Buenos Aires, Instituto de Física de Buenos Aires (IFIBA). Buenos Aires, Argentina}

\author{Muriel Bonetto}
\affiliation{Universidad de Buenos Aires, Facultad de Ciencias Exactas y Naturales, Departamento de Física. Buenos Aires, Argentina.}
\affiliation{CONICET - Universidad de Buenos Aires, Instituto de Física de Buenos Aires (IFIBA). Buenos Aires, Argentina}

\author{Marcelo A. Luda}
\affiliation{Universidad de Buenos Aires, Facultad de Ciencias Exactas y Naturales, Departamento de Física. Buenos Aires, Argentina.}
\affiliation{CEILAP, CITEDEF, Buenos Aires, 1603, Argentina}

\author{Cecilia Cormick}
\affiliation{Instituto de F\'isica Enrique Gaviola, CONICET and Universidad Nacional de C\'ordoba,
Ciudad Universitaria, X5016LAE, C\'ordoba, Argentina}

\author{Christian T. Schmiegelow}
\email{schmiegelow@df.uba.ar}
\affiliation{Universidad de Buenos Aires, Facultad de Ciencias Exactas y Naturales, Departamento de Física. Buenos Aires, Argentina.}
\affiliation{CONICET - Universidad de Buenos Aires, Instituto de Física de Buenos Aires (IFIBA). Buenos Aires, Argentina}

\begin{abstract}

We present an experiment investigating the rotational Doppler effect using a single trapped ion excited by two copropagating vortex laser beams. The setup isolates the azimuthal gradients of the fields, eliminating longitudinal and curvature effects. We provide a detailed characterization of the phenomenon by deterministically positioning a single ion across the beams, achieving a signal which depends on the angular velocity of the ion and the difference of optical orbital angular momentum between the two beams. The interpretation of the measurements is supported by numerical simulations and by a simplified analytical model. Our results reveal key properties of the rotational Doppler effect, showing that it increases approaching the center of the beam and that it is independent of the waist of the beam. This offers insights into the feasibility of super-kicks or super-Doppler shifts for sensing and manipulating atomic motion transverse to the beams' propagation direction.

\end{abstract}

\maketitle

The spatial structure of light beams has been used to sense and change angular momentum in a wide range of systems. The first demonstration of angular momentum transfer from a vortex beam to matter involved making solid micro-sized particles in suspension rotate one way or another by alternating the sign of the beam’s helicity~\cite{he1995direct}. Not long after, the rotation of nano-sized particles in optical tweezers was accomplished in countless experiments and was generally applied as an optical trapping and manipulation technique~\cite{arita2013laser,shen2016trapping}. 
Rotational control of Bose-Einstein condensates using orbital angular momentum of light was also demonstrated~\cite{wright2008optical,lembessis2010light}. At the atomic level, it was shown that optical orbital angular momentum can be transferred to the internal degrees of freedom of single atoms, imposing different selection rules~\cite{schmiegelow2016transfer} and imparting momentum on the atomic center of mass~\cite{stopp2022coherent}. Currently, these techniques are used to improve optical clocks~\cite{lange2022excitation,zanon2023engineering} and quantum gates~\cite{wang2021experimental,vashukevich2022high} as well as for optical trapping of atomic gases and aggregate matter~\cite{gahagan1998trapping,franke2007optical,yang2021optical}.

Vortex beams, such as Laguerre-Gauss beams, have gradients which generate both Doppler shifts~\cite{allen1994azimuthal} and momentum transfer in the direction transverse to the beam's propagation~\cite{allen1992orbital,babiker1994light}.  
That is, they can be used both to sense and to impart motion, just as longitudinal gradients in traveling waves are responsible both for Doppler shifts that depend on the atomic velocity along the propagation direction, and for imparting linear momentum in the same direction. 

A peculiar characteristic of vortex beams is the divergent azimuthal gradient at their center. In particular, the Doppler shift for a Laguerre-Gaussian beam interacting with an atom with velocity $\vec{v}$ gives the usual plane-wave term $\delta_z=kv_z$, an azimuthal term  $\delta_\phi=lv_\phi /r$, and additional curvature terms~\cite{allen1994azimuthal}. Here, $\vec{k}=k\hat z$ is the wavevector, $r$ is the distance of the atom to the beam center, and $l$ is the helicity number, while the subindices in $v$ indicate the corresponding components of the velocity. The azimuthal term $\delta_\phi$, called rotational Doppler effect or azimuthal Doppler shift, is thus  proportional to $l$ and inversely proportional to $r$. 

Two important points can be made about the azimuthal term. First, it only depends on the helicity and the angular velocity $v_\phi/r$ of the ion relative to the beam axis. Hence, it is invariant under changes of the spatial scale for the field alone, as it is independent of the beam waist $\omega_0$ and the wavelength $\lambda$. Second, although it is typically much smaller than the longitudinal term, it can become larger as it diverges at the center of the beam. 

\begin{figure*}[t]
\centering
\includegraphics[]{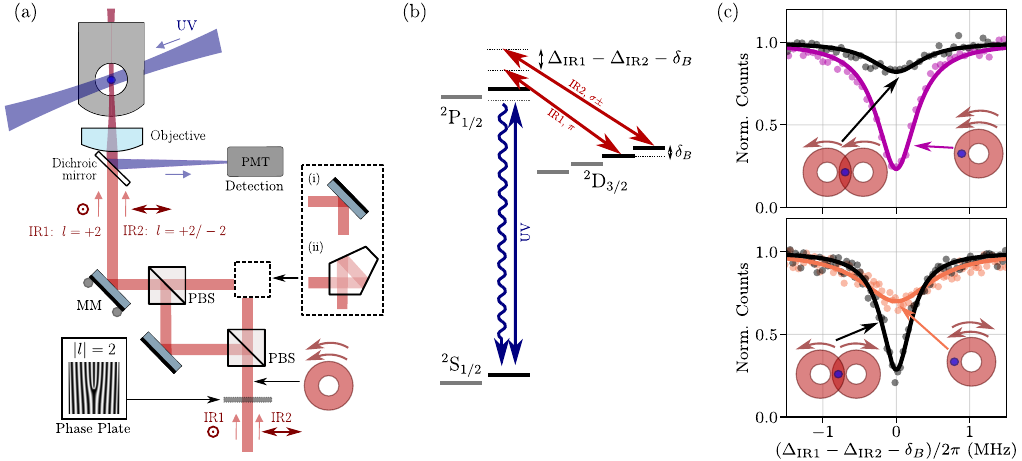}
\caption{
(a) Experimental scheme used to sense the rotational Doppler effect with a single trapped ion (see main text). (b) Relevant lambda-type level scheme. Dark resonances emerge in the atomic spectrum when the difference between the detunings of the IR lasers matches a Zeeman splitting $\delta_B$ between D sublevels. (c) Dark resonance spectra obtained by measuring the UV fluorescence rate while scanning the detuning of one IR beam keeping the other detuning constant. Results shown for different beam configurations as indicated in the panels. 
}
\label{fig:fig1}
\end{figure*}

Berry and Barnett noticed this azimuthal gradient could produce \textit{super-kicks}~\cite{barnett2013superweak}.  That is, upon absorption of a photon, an atom could feel a transverse momentum kick $p_\phi$ which could exceed the standard linear momentum kick along the propagation direction $p_z=\hbar k$. The caveat, however, is that the change of momentum when receiving this transverse kick grows as the atom gets closer to the beam’s center, where the light intensity vanishes. As a result, the probability of observing such super-kicks is very small. It is also difficult to distinguish them from the usual linear momentum kicks~\cite{ivanov2022observability, afanasev2022superkicks}. On the other hand, near the center of the beams one can sense \textit{super-Doppler shifts}, which are expected to diverge as the atom approaches the beam center~\cite{allen1994azimuthal}. Here, too, azimuthal Doppler shifts can surpass the  longitudinal Doppler effect. However, they become harder to detect as one approaches the beam center.

In this paper we present an experiment where we evidence the key properties of the rotational Doppler effect. We use one trapped ion excited by two copropagating beams with equal width and wavevector but possibly different helicities and relative alignments, as shown in Fig.~\ref{fig:fig1}(a). We focus on the analysis of a dark resonance involving interference effects of transitions driven by these two beams. If the beams carry opposite helicities, i.e. $l$ and $-l$, the effects from the longitudinal and the paraxial beam curvature terms are eliminated. This allows us to obtain a signal which only depends on the difference in the azimuthal components of the two beams such that the relative Doppler shift becomes $\delta = 2lv_\phi/r$. 

This idea was previously used to observe the spectrum of a gas at room temperature which was modified due to the super-shift phenomenon~\cite{barreiro2006spectroscopic}. In that case, the observed effects resulted from an integration over all atoms in the gas cell, which continuously moved across the beam.  Here we manage to observe the dependence on location and  direction of the motion by deterministically placing one trapped ion in different positions across the structured beams.  

Our experiments are carried out with a single trapped calcium ion. We use a ring-shaped Paul trap and a standard laser configuration (as described in \cite{nunez2022three}) which allows us to Doppler-cool the ion below mK temperatures. For our secular trap frequencies in the MHz range, this implies a localization below $\sim$100~nm. In this way, a single ion serves as a sub-wavelength probe for the light fields. We set the trap voltages such that the atom is displaced from the trap center. In this situation, the ion is constantly oscillating with a fixed amplitude and direction at the driving frequency of the trap, i.e. it experiences the well-known (and usually spurious) ``micromotion''~\cite{berkeland1998minimization}. In this work, we employ the excess micromotion with controllable direction and amplitude to characterize the rotational Doppler effect. 

We measure, for a single $^{40}$Ca$^+$ ion, the relative Doppler shift between two copropagating, orthogonally polarized, infra-red (IR) lasers near 866~nm, as shown in Fig.~\ref{fig:fig1}(a) and (b). These two IR lasers are blue-detuned by \mbox{$\approx\,30$~MHz} from the dipole transition connecting the metastable $3^2$D$_{3/2}$ and the excited $4^2$P$_{1/2}$ states forming a lambda-type level system. The detunings of these beams ($\Delta_{\mathrm{IR}1} $ and $\Delta_{\mathrm{IR}2} $) can be independently adjusted with a pair of acousto-optic modulators. 

A dark state forms within the D manifold when the difference of detunings matches a difference of Zeeman splittings between D sublevels, $\Delta_{\mathrm{IR}1} -\Delta_{\mathrm{IR}2} =\delta_B$ \cite{suppmat}. We observe this phenomenon through the suppression of ultraviolet (UV) resonance fluorescence corresponding to spontaneous decay from the P manifold to the S ground state. To keep the electron in the P-D manifold and to cool it to mK temperatures, a UV laser near 397~nm red-detuned from the S-P transition by  $\Delta_{\mathrm{UV}} \sim 2\pi \times -20~$MHz is used.  Fluorescence is detected with a photomultiplier tube, while sweeping the frequency of one of the two IR lasers as described in~\cite{nunez2022three} and similar to the Raman configuration used in~\cite{metzner2023two}. 

In a preliminary experiment, the two IR beams are delivered directly to the trap with a Gaussian profile by out-coupling them from a polarization-maintaining single-mode fiber and then re-focusing them onto the ions. To achieve this we bypass the polarization interferometer and the phase plate shown in Fig~\ref{fig:fig1}(a). This approach ensures perfect parallel alignment of the beams, allowing us to observe a linewidth of less than 10~kHz which is roughly 3 orders of magnitude below the Doppler broadening for a temperature of $~1$~mK~\cite{suppmat}. Consequently, our setup demonstrates Doppler-free spectroscopy when the two beams share the same spatial profile and are parallel to each other.

The polarization interferometer shown in Fig.~\ref{fig:fig1}(a) is used to unequivocally determine that we observe a rotational Doppler effect, because it allows us to explore different combinations of beam alignments and helicities. The beams at the output of the fiber are transformed into Laguerre-Gaussian profiles with winding number $|l|=2$ using a diffractive phase plate. Then, the two beams are split in different paths on a Mach-Zehnder-like polarization interferometer built in such a way that one of its arms can have either a mirror or a pentaprism, labeled (i) and (ii) respectively in the figure. The latter adds one extra reflection, so we can make the recombined beams have either equal or opposite helicities by using (i) or (ii). 

When combining two perfectly overlapping beams with the \textit{same helicity} (using the mirror configuration (i)) and placing the ion roughly at the lobe of the beams, we observe the motion has no effect on the dark resonances. This is due to the co-rotating beams canceling out azimuthal Doppler shifts through the two-photon process. This results in a deep dark resonance, shown in purple in the top panel of Fig.~\ref{fig:fig1}(c). 
Next we realign the beams such that the right lobe of one beam coincides with the left lobe of the other beam, and place the ion at this intersection. In this configuration, the ion has opposite azimuthal velocities relative to each beam. The spectrum is now sensitive to the ion's motion, which manifests as a marked reduction in the depth of the dark resonance, as seen in the black dots of the same plot. The reduction of the peak's depth, with a marginal change in its width, is compatible with sensing micromotion on a transition narrower than the modulation frequency~\cite{berkeland1998minimization, suppmat}.

Conversely, when using the pentaprism configuration~(ii) the combined beams have \textit{opposite helicities}. By overlapping them perfectly, we observe a suppression of the dark resonance due to the motion of the ion, shown in orange in the bottom panel of Fig~\ref{fig:fig1}(c). The reduction of the depth of the dip is a consequence of the opposite laser helicities, i.e., a rotational Doppler effect. With the beams realigned such that the ion is placed in the intersection of the lobes, opposite helicities lead to the deep dark resonance shown in black. This is expected, since the opposite angular velocities and helicities lead to the same rotational Doppler shift for the two fields.

\begin{figure*}[htp]
\centering
\includegraphics[]{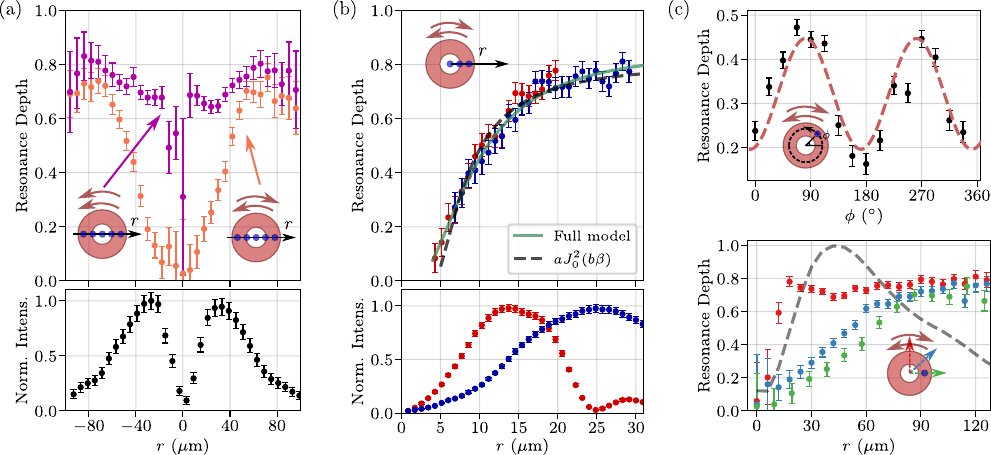}
\caption{
 (a) Comparison of the dark resonance depth for the two collinear configurations by scanning the position of the beams relative to the ion. In the bottom panel we show the beam intensity profile measured through the amount of fluorescence from the ion with only one IR laser on. (b) Comparison of the results for two different choices of the waist of the vortex beams, as shown in the intensity profiles of the bottom panel. In the top panel we show the relative depth of the dark resonance as a function of $r$ varying the power of the beams for each data point such that for all measurements the Rabi frequencies were the same. 
 We plot in green solid line a fit using a full 8-level system simulation, and in gray dashed line a fit using a simplified analytical model, both showing excellent agreement. (c) Measurements of the dependence on the direction of the motion. In the top panel we sweep the position of the beams with respect to the ion in order to vary its angular position $\phi$, keeping $r$ constant. In the bottom panel, we sweep $r$ for three different values of $\phi$. The gray dashed line indicates the intensity profile. 
}
\label{fig:fig2}
\end{figure*}

As a next step, we measure the dependence on the radial position, restricting to well overlapped beams for both the co- and counter-rotating cases (i) and (ii). We control the position of the beams relative to the ion by acting with motorized screws on the last mirror before the focusing lens, indicated MM in Fig.~\ref{fig:fig1}(a). For each position we measure the dark resonance to extract a peak depth and beam intensity by fitting a Lorentzian profile, as in the continuous curves of Fig.~\ref{fig:fig1}(c). From these fits we extract the relative peak depth, shown in the top plot of Fig.~\ref{fig:fig2}(a), whereas the amount of fluorescence with only one IR laser on indicates the spatial dependence of the beam intensity, displayed in the bottom plot of Fig.~\ref{fig:fig2}(a).

For counter-rotating beams there is a clear radial dependence of the peak depth, corresponding to the  orange dots in Fig.~\ref{fig:fig2}(a). In fact, as expected, we see a strong suppression of the dark resonance as one approaches the center, consistent with the divergent relative Doppler shift expected from the expression $\delta=2lv_\phi/r$. As one moves away from the center of the beam, the relative depth values saturate near 0.8, which can be accounted for by off-resonant processes of the multilevel system~\cite{suppmat}. 
In contrast, as seen in the purple dots, for the co-rotating case the peak depth changes much less as one moves across the beam, showing this configuration is almost insensitive to motion irrespective of the position of the ion across the beam. For both cases, one appreciates a large variation in the uncertainty bars, which is a natural consequence of the different beam intensity at each position. 

One of the most intriguing features of the rotational Doppler effect is its scale invariance. To test it, we run a similar sweep as above for two different values of the waist of the vortex beams. The results are shown in Fig.~\ref{fig:fig2}(b), where the top plot displays the relative dip depth and the lower plot the intensity profile for each choice of beam waist. Although we change the waist of the beams by almost a factor of 2, the effect of the rotational Doppler shift on the peak depth is the same in the two cases. 

We note that for the top plot we adjusted the beams' power at each step so that all points were measured with the same Rabi frequencies. This was done to avoid spurious effects due to the intensity-dependent nature of the shape of the dark resonance. This procedure partially limits the range in which the effect can be measured, but it provides data points with similar uncertainties across the whole range measured. 

To compare our measurements with theoretical predictions, we take two approaches: a time-dependent simulation of the 8-level system, and a simplified analytical model of a reduced 3-level system. 
In the first case, we produce a nonlinear-parameter fit where the free parameters are the Rabi frequencies, the relative dephasing between the two IR lasers and the micromotion amplitude. This fit matches accurately the experimental results, as seen in the green solid line of the top panel of Fig.~\ref{fig:fig2}(b). The fit outputs are compatible with our experimental estimates, though the parameter values are highly correlated. More precisely, the data are consistent with azimuthal velocities in the range of 100--200~m$/$s and Rabi frequencies in the $2 \pi\times7$--$9$~MHz range \cite{suppmat}. 

We also developed a simple three-level model that qualitatively describes the observed behavior. We quantify the strength of azimuthal micromotion through the dimensionless modulation parameter $\beta=\delta_\phi^{\rm max}/\Omega_{\mathrm{RF}}$, where $\Omega_{\mathrm{RF}}=2\pi \times 22.135~$MHz is the drive frequency and $\delta_\phi^{\rm max}$ is the maximum azimuthal Doppler shift over an oscillation period. 
Focusing on the weak-pumping regime, for a spectrum which is frequency-modulated by micromotion with $\beta \lesssim 1$, we predict the relative depth of the dark resonance scales with the zero-order Bessel function as $ \mathcal{J}_0^2 (b \beta)$, with a scaling factor $b \approx 2$. This expression is an approximation of a more general formula derived in~\cite{suppmat}. A fit to the data, shown in gray dashed line in Fig.~\ref{fig:fig2}(b), yields a velocity amplitude of $175(5)$~m$/$s, compatible with the interval obtained with the full model. 

Finally, in Fig.~\ref{fig:fig2}(c) we present two plots that highlight the directional character of the motional sensitivity of the rotational Doppler effect. In the top panel of Fig.~\ref{fig:fig2}(c) we azimuthally sweep the position of the beams with respect to the ion. Because the ion micromotion stays fixed in the same direction, this sweep changes the azimuthal projection of the micromotion. For all angles, the distance of the ion to the center of the beams was chosen such that the ion sits at the maximum of the beam profile, in this case $r \sim 42~\mu$m. The depth of the resonance strongly depends on the azimuthal angle, showing two maxima and two minima separated by $180^\circ$. In red dashed line we plot a fit to a sinusoidal curve. The maxima are found at $\phi \simeq 74^\circ, 254^\circ$ which is compatible with the micromotion setting we chose. 

In the bottom panel of Fig. \ref{fig:fig2}(c) we show the radial dependence of the rotational Doppler effect for three different choices of the angular location of the beam. In red, blue and green the measurements are taken along the directions $90^\circ$, $45^\circ$ and $0^\circ$ respectively. The transition between no motional sensitivity to almost disappearance of the dark resonance occurs in different ranges for each curve. This is once more due to the change in the azimuthal projection of the micromotion. We also show in gray dashed line the intensity profile of the beam, to further stress that the spatial dependence of the peak depth is not related to the beam's local intensity.

Our experiments showed sensitivity to velocities of the order of 100~m$/$s for beams with waists of $\sim 15$-$40~\mu$m and distances of down to 5~$\mu$m from the beam's center. To reach the super-Doppler limit, one would need to measure at distances below 1~$\mu$m, where the azimuthal effect becomes bigger than the longitudinal, that is, where $2l/r > 2\pi / \lambda$. In this regime, one would be sensitive to the thermal motion of the ions near the Doppler cooling regime ($\sim$1~mK or $\sim$1~m$/$s) in a similar way as was previously demonstrated with dark resonances~\cite{rossnagel2015fast,tugaye2019absolute}. However, now one would be able to sense the effective temperature transverse to the beam's propagation direction. For these smaller waists, the beams can develop strong longitudinal components at their centers that can influence the sensitivity and must be taken into account. Such effects can be avoided by using azimuthal vector beams or combinations of polarization and helicity which do not generate unwanted fields~\cite{verde2023trapped}.

In this work we used beams with $|l|=2$. The choice of higher helicities increases sensitivity to azimuthal motion, scaling as $\mathcal{S} \propto 2l/r$, but leads to darker beam centers. Indeed, the intensity near the beam axis grows as $I \sim I_0 (r/\omega_0)^{2|l|}$, where $\omega_0$ is the beam waist and $I_0$ is the peak intensity. However, comparing sensitivities at equal intensity $I$ shows that the sensitivity scales as $ \mathcal{S} \propto (I_0/I)^{1/2|l|} \, 2l$. This is typically a growing function of $l$, making the choice of large $l$ more sensitive. We note however that for small $r/\omega_0$ the growth of $\mathcal{S}$ with $l$ is rather slow. 

We conclude by emphasizing that the measurements presented unequivocally demonstrate three key properties of the rotational Doppler shift. Firstly, we evidenced its scale invariance; secondly, we observed increasing sensitivity approaching the center of the beam; and thirdly, we showed the directional character of the effect. This unveils the structural form of the phenomenon, to our knowledge for the first time on a single atom. It also sheds light on the feasibility of observing super-kicks or super-Doppler shifts, which could be used as a motion sensor or actuator in the direction transverse to the propagation direction of the incident beam.

We acknowledge the contribution of M. Drechsler in the building of the ion trap and early experiments. We thank F. Schmidt-Kaler for his continuous support and generosity as well as J. P. Paz, M. A. Latoronda, and A. J. Roncaglia for their unconditional help in the setting up of the laboratory. Finally, we thank U. Poschinger for carefully reading and making suggestions on the manuscript, and M. Ingouville for English proof reading. 
This work was supported by Agencia I+D+i grants PICT2018-3350, PICT2019-4349,  PICT 2020-SERIEA-00959 and PICT 2021-I-A-01288, and Universidad de Buenos Aires grant UBACyT2023-20020220400119BA. The computational resources used in this work were provided (in part) by the HPC center DIRAC, funded by Instituto de Fisica de Buenos Aires (UBA-CONICET) and part of SNCAD-MinCyT initiative, Argentina.

\bibliography{bibliography.bib}

\begin{thebibliography}{28}%
\makeatletter
\providecommand \@ifxundefined [1]{%
 \@ifx{#1\undefined}
}%
\providecommand \@ifnum [1]{%
 \ifnum #1\expandafter \@firstoftwo
 \else \expandafter \@secondoftwo
 \fi
}%
\providecommand \@ifx [1]{%
 \ifx #1\expandafter \@firstoftwo
 \else \expandafter \@secondoftwo
 \fi
}%
\providecommand \natexlab [1]{#1}%
\providecommand \enquote  [1]{``#1''}%
\providecommand \bibnamefont  [1]{#1}%
\providecommand \bibfnamefont [1]{#1}%
\providecommand \citenamefont [1]{#1}%
\providecommand \href@noop [0]{\@secondoftwo}%
\providecommand \href [0]{\begingroup \@sanitize@url \@href}%
\providecommand \@href[1]{\@@startlink{#1}\@@href}%
\providecommand \@@href[1]{\endgroup#1\@@endlink}%
\providecommand \@sanitize@url [0]{\catcode `\\12\catcode `\$12\catcode `\&12\catcode `\#12\catcode `\^12\catcode `\_12\catcode `\%12\relax}%
\providecommand \@@startlink[1]{}%
\providecommand \@@endlink[0]{}%
\providecommand \url  [0]{\begingroup\@sanitize@url \@url }%
\providecommand \@url [1]{\endgroup\@href {#1}{\urlprefix }}%
\providecommand \urlprefix  [0]{URL }%
\providecommand \Eprint [0]{\href }%
\providecommand \doibase [0]{http://dx.doi.org/}%
\providecommand \selectlanguage [0]{\@gobble}%
\providecommand \bibinfo  [0]{\@secondoftwo}%
\providecommand \bibfield  [0]{\@secondoftwo}%
\providecommand \translation [1]{[#1]}%
\providecommand \BibitemOpen [0]{}%
\providecommand \bibitemStop [0]{}%
\providecommand \bibitemNoStop [0]{.\EOS\space}%
\providecommand \EOS [0]{\spacefactor3000\relax}%
\providecommand \BibitemShut  [1]{\csname bibitem#1\endcsname}%
\let\auto@bib@innerbib\@empty
\bibitem [{\citenamefont {He}\ \emph {et~al.}(1995)\citenamefont {He}, \citenamefont {Friese}, \citenamefont {Heckenberg},\ and\ \citenamefont {Rubinsztein-Dunlop}}]{he1995direct}%
  \BibitemOpen
  \bibfield  {author} {\bibinfo {author} {\bibfnamefont {H.}~\bibnamefont {He}}, \bibinfo {author} {\bibfnamefont {M.}~\bibnamefont {Friese}}, \bibinfo {author} {\bibfnamefont {N.}~\bibnamefont {Heckenberg}}, \ and\ \bibinfo {author} {\bibfnamefont {H.}~\bibnamefont {Rubinsztein-Dunlop}},\ }\href {\doibase https://doi.org/10.1103/PhysRevLett.75.826} {\bibfield  {journal} {\bibinfo  {journal} {Physical Review Letters}\ }\textbf {\bibinfo {volume} {75}},\ \bibinfo {pages} {826} (\bibinfo {year} {1995})}\BibitemShut {NoStop}%
\bibitem [{\citenamefont {Arita}\ \emph {et~al.}(2013)\citenamefont {Arita}, \citenamefont {Mazilu},\ and\ \citenamefont {Dholakia}}]{arita2013laser}%
  \BibitemOpen
  \bibfield  {author} {\bibinfo {author} {\bibfnamefont {Y.}~\bibnamefont {Arita}}, \bibinfo {author} {\bibfnamefont {M.}~\bibnamefont {Mazilu}}, \ and\ \bibinfo {author} {\bibfnamefont {K.}~\bibnamefont {Dholakia}},\ }\href {\doibase 10.1038/ncomms3374} {\bibfield  {journal} {\bibinfo  {journal} {Nature Communications}\ }\textbf {\bibinfo {volume} {4}},\ \bibinfo {pages} {2374} (\bibinfo {year} {2013})}\BibitemShut {NoStop}%
\bibitem [{\citenamefont {Shen}\ \emph {et~al.}(2016)\citenamefont {Shen}, \citenamefont {Su}, \citenamefont {Yuan},\ and\ \citenamefont {Shen}}]{shen2016trapping}%
  \BibitemOpen
  \bibfield  {author} {\bibinfo {author} {\bibfnamefont {Z.}~\bibnamefont {Shen}}, \bibinfo {author} {\bibfnamefont {L.}~\bibnamefont {Su}}, \bibinfo {author} {\bibfnamefont {X.-C.}\ \bibnamefont {Yuan}}, \ and\ \bibinfo {author} {\bibfnamefont {Y.-C.}\ \bibnamefont {Shen}},\ }\href {\doibase https://doi.org/10.1063/1.4971981} {\bibfield  {journal} {\bibinfo  {journal} {Applied Physics Letters}\ }\textbf {\bibinfo {volume} {109}},\ \bibinfo {pages} {241901} (\bibinfo {year} {2016})}\BibitemShut {NoStop}%
\bibitem [{\citenamefont {Wright}\ \emph {et~al.}(2008)\citenamefont {Wright}, \citenamefont {Leslie},\ and\ \citenamefont {Bigelow}}]{wright2008optical}%
  \BibitemOpen
  \bibfield  {author} {\bibinfo {author} {\bibfnamefont {K.}~\bibnamefont {Wright}}, \bibinfo {author} {\bibfnamefont {L.}~\bibnamefont {Leslie}}, \ and\ \bibinfo {author} {\bibfnamefont {N.}~\bibnamefont {Bigelow}},\ }\href {\doibase https://doi.org/10.1103/PhysRevA.77.041601} {\bibfield  {journal} {\bibinfo  {journal} {Physical Review A}\ }\textbf {\bibinfo {volume} {77}},\ \bibinfo {pages} {041601} (\bibinfo {year} {2008})}\BibitemShut {NoStop}%
\bibitem [{\citenamefont {Lembessis}\ and\ \citenamefont {Babiker}(2010)}]{lembessis2010light}%
  \BibitemOpen
  \bibfield  {author} {\bibinfo {author} {\bibfnamefont {V.}~\bibnamefont {Lembessis}}\ and\ \bibinfo {author} {\bibfnamefont {M.}~\bibnamefont {Babiker}},\ }\href {\doibase https://doi.org/10.1103/PhysRevA.82.051402} {\bibfield  {journal} {\bibinfo  {journal} {Physical Review A}\ }\textbf {\bibinfo {volume} {82}},\ \bibinfo {pages} {051402} (\bibinfo {year} {2010})}\BibitemShut {NoStop}%
\bibitem [{\citenamefont {Schmiegelow}\ \emph {et~al.}(2016)\citenamefont {Schmiegelow}, \citenamefont {Schulz}, \citenamefont {Kaufmann}, \citenamefont {Ruster}, \citenamefont {Poschinger},\ and\ \citenamefont {Schmidt-Kaler}}]{schmiegelow2016transfer}%
  \BibitemOpen
  \bibfield  {author} {\bibinfo {author} {\bibfnamefont {C.~T.}\ \bibnamefont {Schmiegelow}}, \bibinfo {author} {\bibfnamefont {J.}~\bibnamefont {Schulz}}, \bibinfo {author} {\bibfnamefont {H.}~\bibnamefont {Kaufmann}}, \bibinfo {author} {\bibfnamefont {T.}~\bibnamefont {Ruster}}, \bibinfo {author} {\bibfnamefont {U.~G.}\ \bibnamefont {Poschinger}}, \ and\ \bibinfo {author} {\bibfnamefont {F.}~\bibnamefont {Schmidt-Kaler}},\ }\href {\doibase https://doi.org/10.1038/ncomms12998} {\bibfield  {journal} {\bibinfo  {journal} {Nature Communications}\ }\textbf {\bibinfo {volume} {7}},\ \bibinfo {pages} {12998} (\bibinfo {year} {2016})}\BibitemShut {NoStop}%
\bibitem [{\citenamefont {Stopp}\ \emph {et~al.}(2022)\citenamefont {Stopp}, \citenamefont {Verde}, \citenamefont {Katz}, \citenamefont {Drechsler}, \citenamefont {Schmiegelow},\ and\ \citenamefont {Schmidt-Kaler}}]{stopp2022coherent}%
  \BibitemOpen
  \bibfield  {author} {\bibinfo {author} {\bibfnamefont {F.}~\bibnamefont {Stopp}}, \bibinfo {author} {\bibfnamefont {M.}~\bibnamefont {Verde}}, \bibinfo {author} {\bibfnamefont {M.}~\bibnamefont {Katz}}, \bibinfo {author} {\bibfnamefont {M.}~\bibnamefont {Drechsler}}, \bibinfo {author} {\bibfnamefont {C.~T.}\ \bibnamefont {Schmiegelow}}, \ and\ \bibinfo {author} {\bibfnamefont {F.}~\bibnamefont {Schmidt-Kaler}},\ }\href {\doibase https://doi.org/10.1103/PhysRevLett.129.263603} {\bibfield  {journal} {\bibinfo  {journal} {Physical Review Letters}\ }\textbf {\bibinfo {volume} {129}},\ \bibinfo {pages} {263603} (\bibinfo {year} {2022})}\BibitemShut {NoStop}%
\bibitem [{\citenamefont {Lange}\ \emph {et~al.}(2022)\citenamefont {Lange}, \citenamefont {Huntemann}, \citenamefont {Peshkov}, \citenamefont {Surzhykov},\ and\ \citenamefont {Peik}}]{lange2022excitation}%
  \BibitemOpen
  \bibfield  {author} {\bibinfo {author} {\bibfnamefont {R.}~\bibnamefont {Lange}}, \bibinfo {author} {\bibfnamefont {N.}~\bibnamefont {Huntemann}}, \bibinfo {author} {\bibfnamefont {A.}~\bibnamefont {Peshkov}}, \bibinfo {author} {\bibfnamefont {A.}~\bibnamefont {Surzhykov}}, \ and\ \bibinfo {author} {\bibfnamefont {E.}~\bibnamefont {Peik}},\ }\href {\doibase https://doi.org/10.1103/PhysRevLett.129.253901} {\bibfield  {journal} {\bibinfo  {journal} {Physical Review Letters}\ }\textbf {\bibinfo {volume} {129}},\ \bibinfo {pages} {253901} (\bibinfo {year} {2022})}\BibitemShut {NoStop}%
\bibitem [{\citenamefont {Zanon-Willette}\ \emph {et~al.}(2023)\citenamefont {Zanon-Willette}, \citenamefont {Impens}, \citenamefont {Arimondo}, \citenamefont {Wilkowski}, \citenamefont {Taichenachev},\ and\ \citenamefont {Yudin}}]{zanon2023engineering}%
  \BibitemOpen
  \bibfield  {author} {\bibinfo {author} {\bibfnamefont {T.}~\bibnamefont {Zanon-Willette}}, \bibinfo {author} {\bibfnamefont {F.}~\bibnamefont {Impens}}, \bibinfo {author} {\bibfnamefont {E.}~\bibnamefont {Arimondo}}, \bibinfo {author} {\bibfnamefont {D.}~\bibnamefont {Wilkowski}}, \bibinfo {author} {\bibfnamefont {A.}~\bibnamefont {Taichenachev}}, \ and\ \bibinfo {author} {\bibfnamefont {V.}~\bibnamefont {Yudin}},\ }\href {\doibase https://doi.org/10.1103/PhysRevA.108.043513} {\bibfield  {journal} {\bibinfo  {journal} {Physical Review A}\ }\textbf {\bibinfo {volume} {108}},\ \bibinfo {pages} {043513} (\bibinfo {year} {2023})}\BibitemShut {NoStop}%
\bibitem [{\citenamefont {Wang}\ \emph {et~al.}(2021)\citenamefont {Wang}, \citenamefont {Ru}, \citenamefont {Wang}, \citenamefont {Zhang},\ and\ \citenamefont {Li}}]{wang2021experimental}%
  \BibitemOpen
  \bibfield  {author} {\bibinfo {author} {\bibfnamefont {Y.}~\bibnamefont {Wang}}, \bibinfo {author} {\bibfnamefont {S.}~\bibnamefont {Ru}}, \bibinfo {author} {\bibfnamefont {F.}~\bibnamefont {Wang}}, \bibinfo {author} {\bibfnamefont {P.}~\bibnamefont {Zhang}}, \ and\ \bibinfo {author} {\bibfnamefont {F.}~\bibnamefont {Li}},\ }\href {\doibase 10.1088/2058-9565/ac3c19} {\bibfield  {journal} {\bibinfo  {journal} {Quantum Science and Technology}\ }\textbf {\bibinfo {volume} {7}},\ \bibinfo {pages} {015016} (\bibinfo {year} {2021})}\BibitemShut {NoStop}%
\bibitem [{\citenamefont {Vashukevich}\ \emph {et~al.}(2022)\citenamefont {Vashukevich}, \citenamefont {Bashmakova}, \citenamefont {Golubeva},\ and\ \citenamefont {Golubev}}]{vashukevich2022high}%
  \BibitemOpen
  \bibfield  {author} {\bibinfo {author} {\bibfnamefont {E.}~\bibnamefont {Vashukevich}}, \bibinfo {author} {\bibfnamefont {E.}~\bibnamefont {Bashmakova}}, \bibinfo {author} {\bibfnamefont {T.~Y.}\ \bibnamefont {Golubeva}}, \ and\ \bibinfo {author} {\bibfnamefont {Y.~M.}\ \bibnamefont {Golubev}},\ }\href {\doibase 10.1088/1612-202X/ac45b2} {\bibfield  {journal} {\bibinfo  {journal} {Laser Physics Letters}\ }\textbf {\bibinfo {volume} {19}},\ \bibinfo {pages} {025202} (\bibinfo {year} {2022})}\BibitemShut {NoStop}%
\bibitem [{\citenamefont {Gahagan}\ and\ \citenamefont {Swartzlander}(1998)}]{gahagan1998trapping}%
  \BibitemOpen
  \bibfield  {author} {\bibinfo {author} {\bibfnamefont {K.}~\bibnamefont {Gahagan}}\ and\ \bibinfo {author} {\bibfnamefont {G.}~\bibnamefont {Swartzlander}},\ }\href {\doibase https://doi.org/10.1364/JOSAB.15.000524} {\bibfield  {journal} {\bibinfo  {journal} {Journal of the Optical Society of America B}\ }\textbf {\bibinfo {volume} {15}},\ \bibinfo {pages} {524} (\bibinfo {year} {1998})}\BibitemShut {NoStop}%
\bibitem [{\citenamefont {Franke-Arnold}\ \emph {et~al.}(2007)\citenamefont {Franke-Arnold}, \citenamefont {Leach}, \citenamefont {Padgett}, \citenamefont {Lembessis}, \citenamefont {Ellinas}, \citenamefont {Wright}, \citenamefont {Girkin}, \citenamefont {{\"O}hberg},\ and\ \citenamefont {Arnold}}]{franke2007optical}%
  \BibitemOpen
  \bibfield  {author} {\bibinfo {author} {\bibfnamefont {S.}~\bibnamefont {Franke-Arnold}}, \bibinfo {author} {\bibfnamefont {J.}~\bibnamefont {Leach}}, \bibinfo {author} {\bibfnamefont {M.~J.}\ \bibnamefont {Padgett}}, \bibinfo {author} {\bibfnamefont {V.~E.}\ \bibnamefont {Lembessis}}, \bibinfo {author} {\bibfnamefont {D.}~\bibnamefont {Ellinas}}, \bibinfo {author} {\bibfnamefont {A.~J.}\ \bibnamefont {Wright}}, \bibinfo {author} {\bibfnamefont {J.~M.}\ \bibnamefont {Girkin}}, \bibinfo {author} {\bibfnamefont {P.}~\bibnamefont {{\"O}hberg}}, \ and\ \bibinfo {author} {\bibfnamefont {A.~S.}\ \bibnamefont {Arnold}},\ }\href {\doibase https://doi.org/10.1364/OE.15.008619} {\bibfield  {journal} {\bibinfo  {journal} {Optics Express}\ }\textbf {\bibinfo {volume} {15}},\ \bibinfo {pages} {8619} (\bibinfo {year} {2007})}\BibitemShut {NoStop}%
\bibitem [{\citenamefont {Yang}\ \emph {et~al.}(2021)\citenamefont {Yang}, \citenamefont {Ren}, \citenamefont {Chen}, \citenamefont {Arita},\ and\ \citenamefont {Rosales-Guzm{\'a}n}}]{yang2021optical}%
  \BibitemOpen
  \bibfield  {author} {\bibinfo {author} {\bibfnamefont {Y.}~\bibnamefont {Yang}}, \bibinfo {author} {\bibfnamefont {Y.-X.}\ \bibnamefont {Ren}}, \bibinfo {author} {\bibfnamefont {M.}~\bibnamefont {Chen}}, \bibinfo {author} {\bibfnamefont {Y.}~\bibnamefont {Arita}}, \ and\ \bibinfo {author} {\bibfnamefont {C.}~\bibnamefont {Rosales-Guzm{\'a}n}},\ }\href {\doibase 10.1117/1.AP.3.3.034001} {\bibfield  {journal} {\bibinfo  {journal} {Advanced Photonics}\ }\textbf {\bibinfo {volume} {3}},\ \bibinfo {pages} {034001} (\bibinfo {year} {2021})}\BibitemShut {NoStop}%
\bibitem [{\citenamefont {Allen}\ \emph {et~al.}(1994)\citenamefont {Allen}, \citenamefont {Babiker},\ and\ \citenamefont {Power}}]{allen1994azimuthal}%
  \BibitemOpen
  \bibfield  {author} {\bibinfo {author} {\bibfnamefont {L.}~\bibnamefont {Allen}}, \bibinfo {author} {\bibfnamefont {M.}~\bibnamefont {Babiker}}, \ and\ \bibinfo {author} {\bibfnamefont {W.}~\bibnamefont {Power}},\ }\href {\doibase https://doi.org/10.1016/0030-4018(94)00484-6} {\bibfield  {journal} {\bibinfo  {journal} {Optics Communications}\ }\textbf {\bibinfo {volume} {112}},\ \bibinfo {pages} {141} (\bibinfo {year} {1994})}\BibitemShut {NoStop}%
\bibitem [{\citenamefont {Allen}\ \emph {et~al.}(1992)\citenamefont {Allen}, \citenamefont {Beijersbergen}, \citenamefont {Spreeuw},\ and\ \citenamefont {Woerdman}}]{allen1992orbital}%
  \BibitemOpen
  \bibfield  {author} {\bibinfo {author} {\bibfnamefont {L.}~\bibnamefont {Allen}}, \bibinfo {author} {\bibfnamefont {M.~W.}\ \bibnamefont {Beijersbergen}}, \bibinfo {author} {\bibfnamefont {R.}~\bibnamefont {Spreeuw}}, \ and\ \bibinfo {author} {\bibfnamefont {J.}~\bibnamefont {Woerdman}},\ }\href {\doibase https://doi.org/10.1103/PhysRevA.45.8185} {\bibfield  {journal} {\bibinfo  {journal} {Physical Review A}\ }\textbf {\bibinfo {volume} {45}},\ \bibinfo {pages} {8185} (\bibinfo {year} {1992})}\BibitemShut {NoStop}%
\bibitem [{\citenamefont {Babiker}\ \emph {et~al.}(1994)\citenamefont {Babiker}, \citenamefont {Power},\ and\ \citenamefont {Allen}}]{babiker1994light}%
  \BibitemOpen
  \bibfield  {author} {\bibinfo {author} {\bibfnamefont {M.}~\bibnamefont {Babiker}}, \bibinfo {author} {\bibfnamefont {W.}~\bibnamefont {Power}}, \ and\ \bibinfo {author} {\bibfnamefont {L.}~\bibnamefont {Allen}},\ }\href {\doibase https://doi.org/10.1103/PhysRevLett.73.1239} {\bibfield  {journal} {\bibinfo  {journal} {Physical Review Letters}\ }\textbf {\bibinfo {volume} {73}},\ \bibinfo {pages} {1239} (\bibinfo {year} {1994})}\BibitemShut {NoStop}%
\bibitem [{\citenamefont {Barnett}\ and\ \citenamefont {Berry}(2013)}]{barnett2013superweak}%
  \BibitemOpen
  \bibfield  {author} {\bibinfo {author} {\bibfnamefont {S.~M.}\ \bibnamefont {Barnett}}\ and\ \bibinfo {author} {\bibfnamefont {M.}~\bibnamefont {Berry}},\ }\href {\doibase 10.1088/2040-8978/15/12/125701} {\bibfield  {journal} {\bibinfo  {journal} {Journal of Optics}\ }\textbf {\bibinfo {volume} {15}},\ \bibinfo {pages} {125701} (\bibinfo {year} {2013})}\BibitemShut {NoStop}%
\bibitem [{\citenamefont {Ivanov}\ \emph {et~al.}(2022)\citenamefont {Ivanov}, \citenamefont {Liu},\ and\ \citenamefont {Zhang}}]{ivanov2022observability}%
  \BibitemOpen
  \bibfield  {author} {\bibinfo {author} {\bibfnamefont {I.~P.}\ \bibnamefont {Ivanov}}, \bibinfo {author} {\bibfnamefont {B.}~\bibnamefont {Liu}}, \ and\ \bibinfo {author} {\bibfnamefont {P.}~\bibnamefont {Zhang}},\ }\href {\doibase https://doi.org/10.1103/PhysRevA.105.013522} {\bibfield  {journal} {\bibinfo  {journal} {Physical Review A}\ }\textbf {\bibinfo {volume} {105}},\ \bibinfo {pages} {013522} (\bibinfo {year} {2022})}\BibitemShut {NoStop}%
\bibitem [{\citenamefont {Afanasev}\ \emph {et~al.}(2022)\citenamefont {Afanasev}, \citenamefont {Carlson},\ and\ \citenamefont {Mukherjee}}]{afanasev2022superkicks}%
  \BibitemOpen
  \bibfield  {author} {\bibinfo {author} {\bibfnamefont {A.}~\bibnamefont {Afanasev}}, \bibinfo {author} {\bibfnamefont {C.~E.}\ \bibnamefont {Carlson}}, \ and\ \bibinfo {author} {\bibfnamefont {A.}~\bibnamefont {Mukherjee}},\ }\href {\doibase https://doi.org/10.1103/PhysRevA.105.L061503} {\bibfield  {journal} {\bibinfo  {journal} {Physical Review A}\ }\textbf {\bibinfo {volume} {105}},\ \bibinfo {pages} {L061503} (\bibinfo {year} {2022})}\BibitemShut {NoStop}%
\bibitem [{\citenamefont {Barreiro}\ \emph {et~al.}(2006)\citenamefont {Barreiro}, \citenamefont {Tabosa}, \citenamefont {Failache},\ and\ \citenamefont {Lezama}}]{barreiro2006spectroscopic}%
  \BibitemOpen
  \bibfield  {author} {\bibinfo {author} {\bibfnamefont {S.}~\bibnamefont {Barreiro}}, \bibinfo {author} {\bibfnamefont {J.}~\bibnamefont {Tabosa}}, \bibinfo {author} {\bibfnamefont {H.}~\bibnamefont {Failache}}, \ and\ \bibinfo {author} {\bibfnamefont {A.}~\bibnamefont {Lezama}},\ }\href {\doibase https://doi.org/10.1103/PhysRevLett.97.113601} {\bibfield  {journal} {\bibinfo  {journal} {Physical Review Letters}\ }\textbf {\bibinfo {volume} {97}},\ \bibinfo {pages} {113601} (\bibinfo {year} {2006})}\BibitemShut {NoStop}%
\bibitem [{\citenamefont {Nu{\~n}ez~Barreto}\ \emph {et~al.}(2022)\citenamefont {Nu{\~n}ez~Barreto}, \citenamefont {Drechsler},\ and\ \citenamefont {Schmiegelow}}]{nunez2022three}%
  \BibitemOpen
  \bibfield  {author} {\bibinfo {author} {\bibfnamefont {N.~A.}\ \bibnamefont {Nu{\~n}ez~Barreto}}, \bibinfo {author} {\bibfnamefont {M.}~\bibnamefont {Drechsler}}, \ and\ \bibinfo {author} {\bibfnamefont {C.~T.}\ \bibnamefont {Schmiegelow}},\ }\href {\doibase https://doi.org/10.1103/PhysRevA.106.053708} {\bibfield  {journal} {\bibinfo  {journal} {Physical Review A}\ }\textbf {\bibinfo {volume} {106}},\ \bibinfo {pages} {053708} (\bibinfo {year} {2022})}\BibitemShut {NoStop}%
\bibitem [{\citenamefont {Berkeland}\ \emph {et~al.}(1998)\citenamefont {Berkeland}, \citenamefont {Miller}, \citenamefont {Bergquist}, \citenamefont {Itano},\ and\ \citenamefont {Wineland}}]{berkeland1998minimization}%
  \BibitemOpen
  \bibfield  {author} {\bibinfo {author} {\bibfnamefont {D.}~\bibnamefont {Berkeland}}, \bibinfo {author} {\bibfnamefont {J.}~\bibnamefont {Miller}}, \bibinfo {author} {\bibfnamefont {J.~C.}\ \bibnamefont {Bergquist}}, \bibinfo {author} {\bibfnamefont {W.~M.}\ \bibnamefont {Itano}}, \ and\ \bibinfo {author} {\bibfnamefont {D.~J.}\ \bibnamefont {Wineland}},\ }\href {\doibase https://doi.org/10.1063/1.367318} {\bibfield  {journal} {\bibinfo  {journal} {Journal of Applied Physics}\ }\textbf {\bibinfo {volume} {83}},\ \bibinfo {pages} {5025} (\bibinfo {year} {1998})}\BibitemShut {NoStop}%
\bibitem [{sup()}]{suppmat}%
  \BibitemOpen
  \href@noop {} {}\bibinfo {note} {More details are provided in the supplemental material.}\BibitemShut {Stop}%
\bibitem [{\citenamefont {Metzner}\ \emph {et~al.}(2023)\citenamefont {Metzner}, \citenamefont {Quinn}, \citenamefont {Brudney}, \citenamefont {Moore}, \citenamefont {Burd}, \citenamefont {Wineland},\ and\ \citenamefont {Allcock}}]{metzner2023two}%
  \BibitemOpen
  \bibfield  {author} {\bibinfo {author} {\bibfnamefont {J.}~\bibnamefont {Metzner}}, \bibinfo {author} {\bibfnamefont {A.}~\bibnamefont {Quinn}}, \bibinfo {author} {\bibfnamefont {S.}~\bibnamefont {Brudney}}, \bibinfo {author} {\bibfnamefont {I.}~\bibnamefont {Moore}}, \bibinfo {author} {\bibfnamefont {S.}~\bibnamefont {Burd}}, \bibinfo {author} {\bibfnamefont {D.}~\bibnamefont {Wineland}}, \ and\ \bibinfo {author} {\bibfnamefont {D.}~\bibnamefont {Allcock}},\ }\href@noop {} {\bibfield  {journal} {\bibinfo  {journal} {arXiv preprint arXiv:2312.10847}\ } (\bibinfo {year} {2023})}\BibitemShut {NoStop}%
\bibitem [{\citenamefont {Ro{\ss}nagel}\ \emph {et~al.}(2015)\citenamefont {Ro{\ss}nagel}, \citenamefont {Tolazzi}, \citenamefont {Schmidt-Kaler},\ and\ \citenamefont {Singer}}]{rossnagel2015fast}%
  \BibitemOpen
  \bibfield  {author} {\bibinfo {author} {\bibfnamefont {J.}~\bibnamefont {Ro{\ss}nagel}}, \bibinfo {author} {\bibfnamefont {K.~N.}\ \bibnamefont {Tolazzi}}, \bibinfo {author} {\bibfnamefont {F.}~\bibnamefont {Schmidt-Kaler}}, \ and\ \bibinfo {author} {\bibfnamefont {K.}~\bibnamefont {Singer}},\ }\href {\doibase 10.1088/1367-2630/17/4/045004} {\bibfield  {journal} {\bibinfo  {journal} {New Journal of Physics}\ }\textbf {\bibinfo {volume} {17}},\ \bibinfo {pages} {045004} (\bibinfo {year} {2015})}\BibitemShut {NoStop}%
\bibitem [{\citenamefont {Tugay{\'e}}\ \emph {et~al.}(2019)\citenamefont {Tugay{\'e}}, \citenamefont {Likforman}, \citenamefont {Guibal},\ and\ \citenamefont {Guidoni}}]{tugaye2019absolute}%
  \BibitemOpen
  \bibfield  {author} {\bibinfo {author} {\bibfnamefont {V.}~\bibnamefont {Tugay{\'e}}}, \bibinfo {author} {\bibfnamefont {J.-P.}\ \bibnamefont {Likforman}}, \bibinfo {author} {\bibfnamefont {S.}~\bibnamefont {Guibal}}, \ and\ \bibinfo {author} {\bibfnamefont {L.}~\bibnamefont {Guidoni}},\ }\href {\doibase https://doi.org/10.1103/PhysRevA.99.023412} {\bibfield  {journal} {\bibinfo  {journal} {Physical Review A}\ }\textbf {\bibinfo {volume} {99}},\ \bibinfo {pages} {023412} (\bibinfo {year} {2019})}\BibitemShut {NoStop}%
\bibitem [{\citenamefont {Verde}\ \emph {et~al.}(2023)\citenamefont {Verde}, \citenamefont {Schmiegelow}, \citenamefont {Poschinger},\ and\ \citenamefont {Schmidt-Kaler}}]{verde2023trapped}%
  \BibitemOpen
  \bibfield  {author} {\bibinfo {author} {\bibfnamefont {M.}~\bibnamefont {Verde}}, \bibinfo {author} {\bibfnamefont {C.~T.}\ \bibnamefont {Schmiegelow}}, \bibinfo {author} {\bibfnamefont {U.}~\bibnamefont {Poschinger}}, \ and\ \bibinfo {author} {\bibfnamefont {F.}~\bibnamefont {Schmidt-Kaler}},\ }\href {\doibase https://doi.org/10.1038/s41598-023-48589-1} {\bibfield  {journal} {\bibinfo  {journal} {Scientific Reports}\ }\textbf {\bibinfo {volume} {13}},\ \bibinfo {pages} {21283} (\bibinfo {year} {2023})}\BibitemShut {NoStop}%
\end{thebibliography}%

\end{document}


\title{Supplemental material for \\ ``Observation of Space-Dependent Rotational Doppler Shifts with a Single Ion Probe''}

\author{Nicolás A. Nuñez Barreto}
\affiliation{Universidad de Buenos Aires, Facultad de Ciencias Exactas y Naturales, Departamento de Física. Buenos Aires, Argentina.}
\affiliation{CONICET - Universidad de Buenos Aires, Instituto de Física de Buenos Aires (IFIBA). Buenos Aires, Argentina}

\author{Muriel Bonetto}
\affiliation{Universidad de Buenos Aires, Facultad de Ciencias Exactas y Naturales, Departamento de Física. Buenos Aires, Argentina.}
\affiliation{CONICET - Universidad de Buenos Aires, Instituto de Física de Buenos Aires (IFIBA). Buenos Aires, Argentina}

\author{Marcelo A. Luda}
\affiliation{Universidad de Buenos Aires, Facultad de Ciencias Exactas y Naturales, Departamento de Física. Buenos Aires, Argentina.}
\affiliation{CEILAP, CITEDEF, Buenos Aires, 1603, Argentina}

\author{Cecilia Cormick}
\affiliation{Instituto de F\'isica Enrique Gaviola, CONICET and Universidad Nacional de C\'ordoba,
Ciudad Universitaria, X5016LAE, C\'ordoba, Argentina}

\author{Christian T. Schmiegelow}
\email{schmiegelow@df.uba.ar}
\affiliation{Universidad de Buenos Aires, Facultad de Ciencias Exactas y Naturales, Departamento de Física. Buenos Aires, Argentina.}
\affiliation{CONICET - Universidad de Buenos Aires, Instituto de Física de Buenos Aires (IFIBA). Buenos Aires, Argentina}

\maketitle

In this supplemental material we detail the physical model and the numerical simulations that we used in the main paper to describe the atomic dynamics. Along with this, we explain in detail the spectroscopic technique developed for this work, based on coherent population trapping. Finally, we present the derivation of a simplified theoretical model that qualitatively explains the behavior of the depth of the dark resonance as a function of the radial coordinate.

\section{Model}

\begin{figure}[t]
\centering
\includegraphics[]{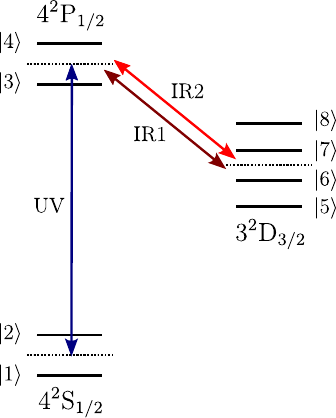}
\caption{
Sketch of the levels of the $^{40}$Ca$^+$ ion relevant for our experiment. The colored arrows indicate the three lasers driving electronic transitions.}
\label{fig:fig1}
\end{figure}

We begin by describing the atomic dynamics relevant for our experiment. We consider the 8-level system of the $^{40}$Ca$^+$ ion interacting with three lasers as depicted in Fig.~\ref{fig:fig1}. A magnetic field $ B\sim 4~\mathrm{G}$ is used to induce Zeeman splittings of the order of 10 MHz. Two IR lasers near 866~nm, labeled IR1 and IR2, are used to drive transitions between the $3^2$D$_{3/2}$ metastable manifold and the upper $4^2$P$_{1/2}$ states. Both IR lasers have linear polarizations. The IR2 laser is polarized orthogonal to $\vec{B}$, i.e. $\sigma_+ + \sigma_-$ polarized, and depopulates all D states. The polarization of the IR1 laser is parallel to $\vec{B}$, i.e. $\pi$ polarized, and this is the laser whose detuning we sweep in order to obtain the spectra. 
Additionally, the $4^2$P$_{1/2}$ states decay via a dipolar transition to the ground $4^2$S$_{1/2}$ states. To keep the cycle closed and allow for Doppler cooling and fluorescence observation, we drive this transition with an ultraviolet (UV) laser near 397~nm that is chosen to be $\sigma_+ + \sigma_-$ polarized. The UV laser can be treated as a plane wave, whereas the IR lasers are vortex beams.

We describe the dynamics with a master equation for the density matrix $\rho$ of the electronic degrees of freedom as
\begin{equation}
    \frac{d\rho}{dt} = - \frac{i} {\hbar} [\mathcal{H},\rho] + \mathcal{L}(\rho)\, .
\end{equation}
Here, $\mathcal{H}$ is the Hamiltonian of the 8 levels interacting with the three lasers fields using a rotating-wave approximation and in a frame rotating with the frequencies of the UV and one of the IR lasers (which we choose to be IR2). Since we have two IR lasers that address the same transitions, the Hamiltonian for an atom at rest will be explicitly time-dependent and periodic with a frequency equal to the difference of detunings $\Delta_{\mathrm{IR2}}-\Delta_{\mathrm{IR1}}$ between the two IR lasers.
The incoherent dynamics related to spontaneous decays from the upper levels to the lower ones, and dephasing effects due to finite laser linewidths and temperature of the ion, are modeled via the superoperator $\mathcal{L}(\rho)$.

In addition, we consider micromotion in the dynamics. This is included by modulating the detuning $\Delta$ of each laser in the form $\Delta'(t) = \Delta + \delta \cos(\Omega_{\mathrm{RF}}t)$. This adds another periodic time-dependence in the Hamiltonian, at the drive frequency $\Omega_{\mathrm{RF}}$. The prefactor $\delta$ is the scalar product between the gradient of the phase of the laser's electric field and the velocity amplitude $\vec v$ of the micromotion. This contribution to the dynamics is due to the Doppler shift. For plane waves, the shift takes the form $\delta = - \vec{k} \cdot \vec{v} = k v_z$, where $\vec{k}=k \hat{z}$ is the wavevector of the laser. For Laguerre-Gaussian (LG) beams with radial number $p$ and winding number $l$, the expression of the shift in cylindrical coordinates $\vec{r}=(r,\phi,z)$ is
\begin{equation}
\begin{aligned}
\delta_{\mathrm{LG}} = {} & -\Biggl[k + \frac{kr^2}{2(z^2+z_R^2)}\left( \frac{2z^2}{z^2+z_R^2}-1 \right) \\
& - \frac{(2p+|l|+1)z_R}{z^2+z_R^2}\Biggl]v_z  - \left( \frac{krz}{z^2+z_R^2} \right)v_r - \left( \frac{l}{r} \right)v_{\phi}.
\end{aligned}
\end{equation}
Here, $z_R=\frac{1}{2}k \omega_0$ is the Rayleigh range of the beam, with $\omega_0$ the beam waist, and $v_z$, $v_r$, $v_\phi$ are the longitudinal, radial and azimuthal components, respectively, of the velocity of the ion with respect to the propagation direction of the laser. 

From this formula we can appreciate that there are contributions from the three components of the atom's velocity. Particularly, the plane-wave phase is still present in the longitudinal component and is usually the dominating term. There are also two extra longitudinal terms, one given to the Gaussian curvature of the beam and the other one from the Gouy phase of the LG beam. The radial contribution also results from the Gaussian beam's curvature. Finally, the term that depends on the angular velocity is known as azimuthal Doppler shift or rotational Doppler effect. Our setup is designed in such a way that the contributions to $\delta_{\rm LG}$ involving two-photon processes driving the D manifold cancel out, with the exception of the azimuthal term.

The resulting time-dependent master equation is cast in the form of $8 \times 8$ differential equations for the elements of $\rho$ in the Zeeman basis. Considering that $\rho^\dagger = \rho$ and $\Tr \rho = 1$, we are left with 35 linearly independent equations (for generally complex matrix elements). To obtain the atomic spectra we perform a numerical solution of the differential equations given an initial condition where the atomic population is all in one of the S states. 
We evolve the system for $30~\mu$s to eliminate the initial transient and then average the electronic populations over $70~\mu$s.

To numerically calculate the spectra, we repeat the simulation for different values of the IR1 detuning $\Delta_{\mathrm{IR1}}$ and compute the sum of the populations of the two excited states, since this quantity will be proportional to the scattered fluorescence. In Fig.~\ref{fig:fig2}(a) we show an example of a simulated spectrum. The fluorescence exhibits several dips, commonly called dark resonances, which we explain in detail in the next section.

\section{Dark resonances}

\begin{figure*}[t]
\centering
\includegraphics[]{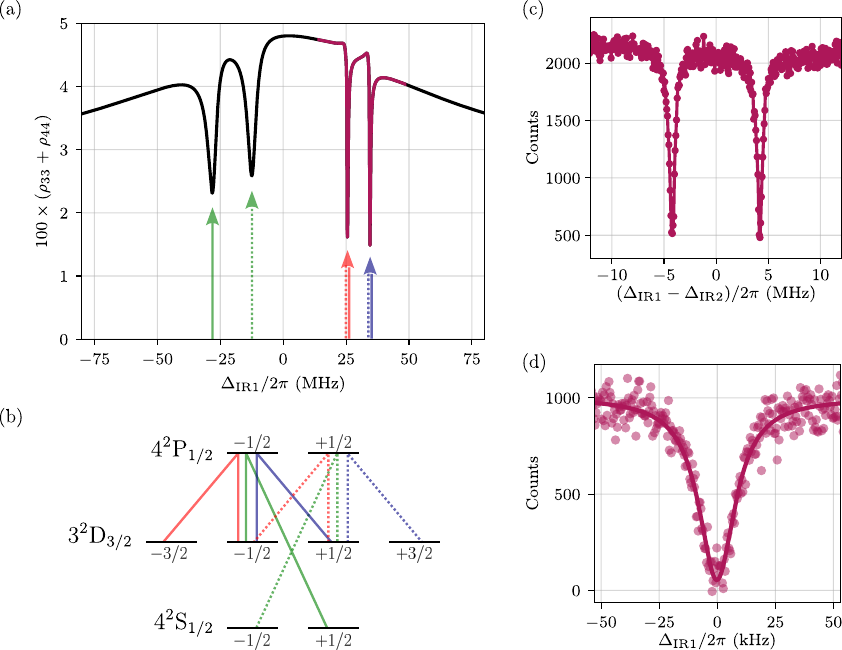}
\caption{
Dark resonances emerging from sweeping the IR1 detuning around resonance. (a) Example spectrum where two pairs of dips manifest. (b) Level scheme indicating the different dark resonances. Vertical lines correspond to transitions driven by $\pi$ polarized lasers, while oblique transitions are driven by $\sigma_+ + \sigma_-$ polarized lasers. For our configuration, six different dark states can form. They manifest as dips in the fluorescence, indicated in the spectrum. (c) Experimental observation of the dark resonances involving D sublevels only. (d) Lorentzian fit of one of the dark resonances of (c), after a major reduction of the beams intensities to decrease power broadening. 
}
\label{fig:fig2}
\end{figure*}

In any lambda-type level system driven by two lasers there can be interference between the two transition amplitudes. This happens when the relative detunings of the two lasers with respect to the upper level match the energy difference of the lower states, creating a dark state. This manifests in the atomic spectrum as a dark resonance. In our system, there are many possible three-level subsystems allowing for several different dark resonances. Their precise number and location will depend on the relation between the polarizations of the lasers with respect to the direction of the magnetic field, and on the frequency detuning of each of the three lasers. In this section, we will not include Doppler shifts in order to focus on the origin and features of the dark resonances.

In Fig.~\ref{fig:fig2}(a) and (b) we indicate all the dark resonances that emerge for the polarizations of our experiment. We use lines of different colors and/or styles for the combinations of S-D or D-D states that give rise to each dark resonance. In Fig.~\ref{fig:fig2}(a) we show a simulation of the corresponding spectrum, where the respective dark resonances emerge, marked with the same code. In the left part of the plot two S-D resonances can be seen, centered at $\Delta_{\mathrm{IR1}}=\Delta_{\mathrm{UV}} = 2\pi \times -20~$MHz. In the right, the two resonances, centered at $\Delta_{\mathrm{IR1}}=\Delta_{\mathrm{IR2}} = 2\pi \times 30~$MHz, are due to D-D superpositions; in this case there are two pairs of superpositions corresponding to each fluorescence dip. One can note that these last resonances are much narrower than the first two. This is because the two IR lasers share the same source while the IR and the UV are from different lasers. Therefore, a dephasing of 100~kHz is included for the UV-IR lasers while no dephasing is considered between the IR lasers. In this case, the resonance width of the D-D superposition is only due to power broadening. This spectrum was simulated with parameters that are similar to the ones used in the experiment, i.e., low saturation in the UV laser and high saturation in the two IR lasers. This generates a power broadening of the overall spectrum. Furthermore, for $\Delta_{\mathrm{IR1}}\rightarrow \pm \infty$, the excited states populations do not tend to zero, since the IR2 laser makes a closed fluorescence cycle with the UV laser.

It can be noticed in Fig.~\ref{fig:fig2}(b) that each combination of D states that leads to a dark resonance involves one sublevel that is also off-resonantly pumped into the P manifold. This makes the depth of the dark resonances strongly dependent on the Rabi frequencies of the IR lasers: if they are much smaller than the Zeeman splitting, then off-resonant processes can be neglected and the dark resonance is almost perfect. Otherwise, there are no strictly dark states since they are depleted by off-resonant processes, so that the fluorescence can exhibit a dip with a minimum clearly away from zero. 

In Fig.~\ref{fig:fig2}(c) we show a measurement of the two D-D resonances. This measurement was carried out in a configuration that minimized motional effects, i.e. insensitive to relative Doppler shifts between the two IR lasers. We also show in Fig.~\ref{fig:fig2}(d) a zoom of the left resonance after reducing drastically the intensity of both IR lasers. We can appreciate that it can reach a width of less than 10~kHz, well below the Doppler broadening for a temperature of $~1$~mK. Furthermore, the dark resonance in this plot is almost perfect, with its minimum very close to zero fluorescence. 

In the set of experiments of the main text, we use excess micromotion driven purposely to induce an oscillation of the ion allowing us to sense the rotational Doppler effect. This causes a reduction in the depth of the dark resonance due to atomic motion, when the laser configuration is chosen to be sensitive to it. Thermal motion is also present and can have an effect in the shape of the dark resonance. The different nature of each kind of dynamics gives rise to different manifestations in the spectrum. 

Micromotion is a driven oscillation with fixed amplitude and direction, at a trap's drive frequency of $
\sim$ 22 MHz. As this frequency is substantially bigger than all the dark resonance linewidths we work with (typically in the MHz regime) we expect it to generate a resolved frequency modulated spectra. That is, we expect micromotion to generate a decrease in the depth of the dark resonance, while keeping its width constant. 

\begin{figure}[t]
\centering
\includegraphics[]{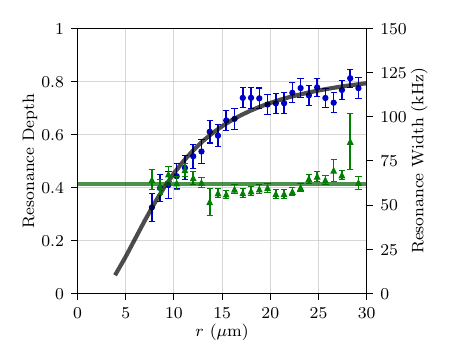}
\caption{
Characteristics of the dark resonance varying $r$, for the measurements of Fig.~2(a) of the main paper. In green, we show the width of the dark resonance for each value of $r$, along with a horizontal line representing the mean value of the widths. In blue we display the corresponding value of the depth of the dark resonance, with the fit to the full model described in Sec. \ref{sec3}.}
\label{fig:fig3}
\end{figure}

To check this in Fig.~\ref{fig:fig3}, we plot the dark resonance depth (blue) alongside the resonance width (green), as the ion is moved towards the center of the beam (the data shown corresponds to the same measurement as in Fig.~2(a) of the main paper). For each point, a scan across the resonance was measured. The widths and depths were then extracted from the spectra by a Lorentzian fit. As can be seen, the widths obtained are approximately constant along the scan while the depth decreases as one approaches the center of the beam. 

Moreover, as a double check, we find that under certain conditions, we are able to observe the micromotion echoes of the dark resonances. These are, however, tricky to see as both the convolution of the dark resonance and the actual natural resonance play a role. Therefore, the region of parameters for the intensities of the lasers and for the amplitude of the micromotion where the echoes are clearly resolved is quite narrow.

\section{Fit of the experimental data with the full numerical model}
\label{sec3}

In Fig.~2 of the main paper we performed a fit to the measured data using the model described in the previous section. To compare the experiments with the theory, we computed the depth of each dark resonance measured, defined as $d=1-R$, where $R$ is the ratio between the intensities of the fluorescence at the resonance and sufficiently away from it, after subtracting the corresponding backgrounds. To reduce uncertainties, we obtained those fluorescence values by fitting the resonance with a Lorentzian function.

We perform a nonlinear-parameter fit by numerically solving the differential equations of the system as described before. The Rabi frequencies of the three lasers, the relative dephasing between the D levels and the ion's velocity are left as free parameters. We obtain values which are compatible with our experimental estimates, though highly correlated. Nevertheless, we stress that the goal of this fit is not the determination of these free parameters. Rather, the point is to confirm that our model reproduces adequately the  experimental results in Fig.~2 of the main paper.

\begin{figure}[t]
\centering
\includegraphics[]{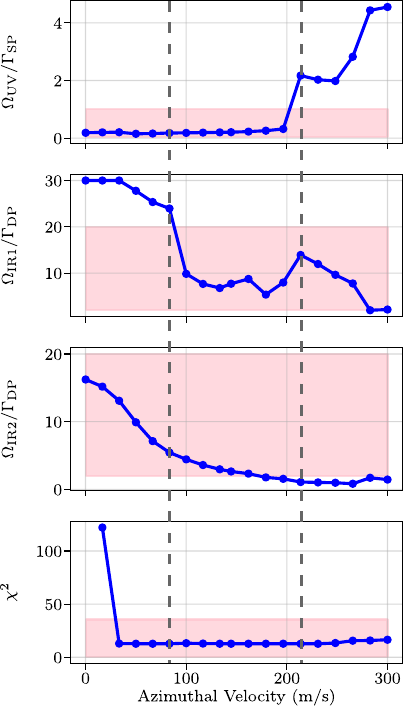}
\caption{
Range of parameters obtained from fitting the experimental data with the full model. The zone marked between dashed lines corresponds to the parameters space in which the Rabi frequencies are in good agreement with estimations.}
\label{fig:fig4}
\end{figure}

To estimate a range of parameters compatible with the experimental results we fix the velocity amplitude to different values and find the best fit for the remaining four free parameters. With $31$ data points, this results in a fit with $\nu = 27$ degrees of freedom. To determine an interval for the possible velocities we follow two criteria. Initially we consider only the velocity values for which the $\chi^2$ of the fit is lower than the 90th percentile of the corresponding $\chi^2_{(\nu)}$ distribution ($\chi^2 < 36.7$ for $\nu = 27$). We then restrict this interval to those where the parameters that result from the fit are consistent with our experimental values. This criteria are indicated by the pink regions in Fig.~\ref{fig:fig4} and the estimated interval for the velocity that follows from the analysis is marked by vertical dashed lines. This figure was produced setting $\Delta_\mathrm{UV} = 2 \pi \times -30$~MHz, $\Delta_\mathrm{IR1} = 2 \pi \times 40$~MHz. Modifying these values results in velocity estimates varying by less than 20~m$/$s, and this specific set of detunings is chosen as they yield the most extended intervals among those examined. The remaining fit parameter, the dephasing, remains below $1$~kHz for all velocities.

\section{Simplified analytical model for the depth of the dark resonances}

In this section we develop a much simpler model to qualitatively understand the dynamics of interest. To facilitate the analytical calculations, we restrict to only three levels, which constitute the smallest system that can display such behavior. We consider an excited state $\ket{e}$ that represents a single P level. We include also two stable levels $\ket{1}$ and $\ket{2}$ that are dipole-coupled to the excited level, and that represent two D sublevels. Each of the transitions is driven by a laser with opposite Doppler shift when the ion moves due to the micromotion. The decay from P to S levels, followed by repumping by the UV laser, is replaced by an effective dephasing of the P with respect to the D states, at a large rate associated with the upper state linewidth (which is further broadened by the UV drive). 

The relevant parameters are: the effective dephasing rate $\gamma \gtrsim \Gamma/2$, the Rabi frequency $\Omega$ for each of the two driven transitions (we assume equal Rabi frequencies for simplicity), the frequency $\Omega_{\rm RF}$ of the micromotion, the detuning $\Delta_j$ of each laser, with $j=1,2$, and the decay rate $\Gamma_D$ to the stable manifold. We assume that the decay to each stable sublevel is equally likely. For this calculation, we assume that the S-P pumping is strong, but the D-P pumping is weak, $\Omega\ll\Gamma_D$. Finally, we include the effect of off-resonant processes as a depolarizing noise within the stable subspace, acting at a very small rate $\gamma' \ll \Gamma_D$. We also use $\Gamma_D\ll\gamma$, an approximation which is not essential but is helpful to obtain compact formulas.

In the atomic reference frame, the Hamiltonian of this system is given by
\begin{multline}
 \frac{\mathcal{H}(t)}{\hbar} = \frac{\Omega}{2} \sum_{n=-\infty}^\infty e^{int\Omega_{\rm RF} } i^n J_n(\beta) \ket{e} \\\Big[\, \bra{1} e^{-i\Delta_1 t}
 + (-1)^n \bra{2} e^{-i\Delta_2 t} \Big] + {\rm H. c.}
\end{multline}
Here, $\beta=\delta_\phi^{\rm max}/\Omega_{\mathrm{RF}}$ is the micromotion parameter, and we assume that each of the transitions experiences Doppler shifts with opposite signs, so that we can use the parity of the Bessel functions, $J_n (-\beta) = (-1)^n J_n (\beta)$.

Given our choice of equal Rabi frequencies for the two transitions, dark resonances can be best analyzed using the basis
\begin{equation}
 \ket{\pm} = \frac{1}{\sqrt{2}} (\ket{1} \pm \ket{2} ) \,.
\end{equation}
The above Hamiltonian can then be rewritten as
\begin{multline}
 \frac{\mathcal{H}(t)}{\hbar} = \frac{\Omega}{2\sqrt{2}} \sum_n e^{int\Omega_{\rm RF} } i^n J_n(\beta) \ket{e} \\\Big\{\, \bra{+} \Big[e^{-i\Delta_1 t} + (-1)^n e^{-i\Delta_2 t} \Big]\\
 + \bra{-} \Big[e^{-i\Delta_1 t} - (-1)^n e^{-i\Delta_2 t} \Big] \Big\} + {\rm H. c.}
\end{multline}

In particular, at the resonance point $\Delta_1=\Delta_2=\Delta$, we obtain
\begin{multline}
 \frac{\mathcal{H}(t)}{\hbar} = \frac{\Omega}{2\sqrt{2}} \sum_n e^{it(n\Omega_{\rm RF}-\Delta)}\, i^n J_n(\beta) \ket{e} \\\{\, \bra{+} [1 + (-1)^n ] 
 + \bra{-} [1 - (-1)^n ] \} + {\rm H. c.}
\end{multline}
In absence of micromotion, $\beta=0$ and all Bessel functions vanish except for the case $n=0$. This means that the odd combination of stable sublevels is decoupled from the laser fields and the asymptotic state for this system is the dark state $\ket{-}$. More generally, all even values of $n$ are associated with pumping from state $\ket{+}$, and all odd $n$ indices correspond to pumping from $\ket{-}$. A dark state appears when one of these states experiences no pumping.

We now proceed to estimate the depth of this dark resonance. For this we resort to rate equations, which means we focus on the evolution of the diagonal elements of the density matrix only. In order to calculate the fluorescence at the resonance point, we will use rate equations in the basis $\{\ket{e}, \ket{+}, \ket{-}\}$. On the other hand, sufficiently far from the resonance we will use the basis $\{\ket{e}, \ket{1}, \ket{2}\}$. We note that generally the width of the dark resonance, and therefore what ``sufficiently far'' means, depends on the Rabi frequency $\Omega$.

We tackle first the case on resonance. The state is then described by the populations $p_e$ and $p_\pm$, which evolve according to
\begin{eqnarray}
 \dot p_e &=& - \Gamma_D\, p_e + \gamma_{\rm pump}^+\, p_+ + \gamma_{\rm pump}^-\, p_-\\
 \dot p_\pm &=& \frac{\Gamma_D}{2}\, p_e - \gamma_{\rm pump}^\pm\, p_\pm + \gamma' (p_\mp - p_\pm),
\end{eqnarray}
where $\gamma_{ \rm pump}^\pm$ are total pumping rates from the states $\ket{\pm}$ into $\ket{e}$.
Since we are assuming that the pumping is very weak, we can approximate
\begin{equation}
 \gamma_{\rm pump}^\pm = \frac{\gamma}{2} \sum_{n=-\infty}^\infty \frac{(\Omega_n^\pm)^2}{\gamma^2 + (\Delta_n^\pm)^2},
\end{equation}
where the values of $\Omega_n^\pm$ and $\Delta_n^\pm$ can be extracted from the Hamiltonian for $\Delta_1=\Delta_2=\Delta$. In particular, $\Omega_n^+$ vanishes for odd $n$, and $\Omega_n^-$ for even $n$.
For sufficiently weak pumping, the equilibrium value of the excited state population is given by
\begin{equation}
 p_e \simeq \frac{2}{\Gamma_D} \,\frac{\gamma_{\rm pump}^+ \, \gamma_{\rm pump}^- + \gamma' (\gamma_{\rm pump}^++\gamma_{\rm pump}^-)}{\gamma_{\rm pump}^++\gamma_{\rm pump}^- + 4 \gamma'}.
\end{equation}

The case out of resonance can be carried out in the same form but in the basis $\{\ket{e}, \ket{1}, \ket{2}\}$. Putting all ingredients together, we obtain that the depth of the dark resonances can be written as
\begin{equation}
d(\beta) \simeq 1 - \frac{4}{F_+ + F_-}\, \frac{ F_+ \, F_- + \tilde\gamma (F_+ + F_-) }{F_+ + F_-+4\tilde\gamma},
\label{eq:model1}
\end{equation}
with
\begin{equation}
 \tilde \gamma = \frac{\gamma' \gamma}{\Omega^2}
\end{equation}
and
\begin{equation}
F_\pm \simeq \sum_n \frac{J_n^2(\beta)}{ 1 + \left(\dfrac{\Delta - n \Omega_{\rm RF}}{\gamma}\right)^2} \,.
\end{equation}
In this last expression the sum runs over even (odd) indices for the $+$ ($-$) sign. In order to achieve the result in Eq.~\eqref{eq:model1}, we assumed that the dark resonances are narrow enough so we can evaluate the depth of the resonance by shifting the value of $\Delta_2-\Delta_1$ in an amount which is still much smaller than the typical scale of $\Delta$ and $\Omega_{\rm RF}$.

The depth of the dark resonance for $\beta=0$ takes the form
\begin{equation}
 d(\beta=0) = \frac{1}{1+ 4 \tilde \gamma \left[1+\left(\dfrac{\Delta}{\gamma}\right)^2\right]} 
\end{equation}
so that it is equal to 1 in absence of noise within the stable manifold.

We note that for values of $\Delta$, $\Omega_{\rm RF}$ and $\gamma$ comparable to the ones in our experiment, in the regime $\beta < 1.25$ the function $d(\beta)$ can be well approximated by a horizontal rescaling of only one Bessel function,
\begin{equation}
 d(\beta)\simeq a J_0^2( b \beta)
\label{eq:model2}
\end{equation}
with $b \simeq 2$. This approximation, however, does not hold for larger values of $\beta$ or for arbitrary parameter ranges.

In Fig. 2(b) of the main paper we show a fit to the experimental data using Eq.~\eqref{eq:model2}. The fitted values for $r$ satisfy the condition $\beta<1$, so the approximation made is valid. For this fitting, we fixed $b=2$ and retrieved the azimuthal velocity of the ion, reported in the paper.
